\def\gcmm3{{\,{\rm g\,cm^{-3}}}}

\def\la{\mathrel{\mathpalette\fun <}}
\def\ga{\mathrel{\mathpalette\fun >}}
\def\fun#1#2{\lower3.6pt\vbox{\baselineskip0pt\lineskip.9pt
  \ialign{$\mathsurround=0pt#1\hfil##\hfil$\crcr#2\crcr\sim\crcr}}}
\documentstyle[preprint,prd,aps,epsfig,floats,graphics]{revtex}
\draft

\begin{document}

\tighten

\title{
\vskip1.55cm
Top-Down Models and Extremely\\
High Energy Cosmic Rays}

\author{O.E. Kalashev, V.A. Kuzmin, D.V. Semikoz}
\address{Institute for Nuclear Research of Russian Academy of Sciences,
60th October Anniversary Prosp. 7a, Moscow 117312, Russia\\
E-mails: kalashev@ms2.inr.ac.ru, kuzmin@ms2.inr.ac.ru, semikoz@ms2.inr.ac.ru}

\maketitle

\begin{abstract}
{\it
We developed numerical code for calculation of the extragalactic component
of the spectra of leptons, nucleons and $\gamma$-rays resulting from
``top-down'' (non-acceleration) models for the case of uniform and
isotropic source distribution.
We explored two different classes of ``top-down'' scenarios:
the wide earlier investigated class of X particles coming from collapse and/or
annihilation of cosmic topological defects (such as cosmic strings, monopoles,
etc.) and the models of super-heavy long-living X particles with lifetime
of the order or much greater than the current Universe age.
}

\end{abstract}

\section{Introduction}
The ultra-high energy cosmic ray (UHECR) events observed above
$10^{20}$ eV~\cite{fe1,agasa1} are difficult to explain
within conventional models involving first order Fermi acceleration of
charged particles at astrophysical shocks~\cite{Blandford}. It is hard
to accelerate protons and heavy nuclei up to such energies even in the
most powerful astrophysical objects~\cite{Hillas} such as radio
galaxies and active galactic nuclei. Nucleons of the energies above
$\simeq4\times10^{19}$eV are strongly decelerated due to photo-pion
production on the cosmic microwave background (CMB) --- the
Greisen-Zatsepin-Kuzmin (GZK) effect~\cite{GZK} --- on the distances
less than $\simeq100\,$Mpc~\cite{SSB}. Ultra high energy electrons
lose their energy even faster due to synchrotron radiation in
galactic as well as extragalactic magnetic field. Heavy nuclei are
photodisintegrated in the CMB within a few Mpc~\cite{Puget}.
There are no evident astronomical sources of UHECR within $\simeq100$~Mpc
of the Earth in the frame of conventional acceleration models.

To avoid these difficulties one may assume that UHECR are
created directly at energies comparable to or exceeding the observed
ones rather than being accelerated from lower energies. In these
so-called ``top-down'' scenarios $\gamma$-rays, leptons and nucleons
are initially produced at ultra-high energies by the decays of
supermassive particles generically called X-particles. The
sources of X-particles could be topological defects (TD) such as
cosmic strings or magnetic monopoles that could be produced in the
early Universe during symmetry-breaking phase transitions envisaged in
Grand Unified Theories. In the inflationary Universe, the relevant TD
could be formed in a phase transition at the end of inflation.
Another possibility is to produce super-heavy long-living particles
thermally during reheating epoch of the Universe~\cite{KuzRub} or from
vacuum fluctuations during inflation \cite{KuzTkach}.

There were several papers dedicated to the calculation of the
observable spectra of UHECR~\cite{Lee,Sigl98,Berezinsky98}.
The most
detailed discussion of propagation of nucleons, photons and electrons
is given in ref.~\cite{Lee}.  It has, however, some drawbacks which
we discuss below. We have developed an independent numerical
code for calculation of the extragalactic component of the spectra of
leptons, nucleons and $\gamma$-rays resulting from an X-particle
decays in the top-down models of UHECR, for the case of uniform and
isotropic source distribution. Our general strategy is similar to that
of ref.~\cite{Lee}, but we use another numerical scheme for the
solution of transport equations, and take into account process of
proton-photon scattering $p\gamma_b\rightarrow p\gamma$ disregarded
previously.

In this paper we present the results of numerical calculations of the
observable spectra in two top-down models. The first one belongs
to the class of models examined in \cite{Lee}, and we use it for
testing our code. The second model is based on the decays of
superheavy relic particles; it was not considered in ref.~\cite{Lee}
We shall discuss the differences in the observable spectra predicted
in different models, and derive constraint on X-particle lifetime in
the second model.

\section{Injection Spectra and evolution}

The generic top-down mechanism of production of UHECR looks as
follows: the X-particles with typical GUT masses $m_X$ of the order of
$10^{14}-10^{16}$~GeV decay into leptons and quarks. The strongly
interacting quarks fragment into jets of hadrons resulting in mesons
and baryons. Mesons, in turn, decay into photons, electrons and
neutrino. Together with baryons these particles form the primary
injection spectrum.

Injection spectrum for a given particle species $(a)$ from the
hadronic channel can be written as
\begin{equation}
\Phi_a(E,t)={dn_{d,X}(t)\over dt}{n_q+n_e\over m_X}{dN_a(x)\over dx}\,,\label{injh}
\end{equation}
where $x\equiv E(n_q+n_e)/m_X$, and $dN_a/dx$ is the effective
fragmentation function describing the production of the particles of
species $(a)$ by the original quark.  The {\it total\/} hadronic
fragmentation spectrum $dN_h/dx$ in the energy range of our interest
is not yet known definitely.  Some possibilities were described in
ref.  \cite{Sigl98}. In our current calculations  we have used the
following fragmentation function, 
\begin{equation}
{dN_h(x)\over dx}=\left\{\begin{array}{ll}
             {15\over16}x^{-1.5}(1-x)^2 & \mbox{if $x_0\leq x\leq1$}\\
             0 & \mbox{otherwise}\end{array}\right. \,,\label{frag}
\end{equation}
where the value $x_0$ corresponds to a cut-off energy $\sim1{\,{\rm
GeV}}$. Equation (\ref{frag}) is one of the simplest QCD-motivated
forms of hadronic fragmentation spectrum~\cite{Hill}.  Assuming a
nucleon content of $\simeq3\%$ and the rest equally distributed among
the three types of pions, we can write the fragmentation spectra
as~\cite{bhs,abs}
\begin{eqnarray}
{dN_N(x)\over dx}&=&(0.03)\,{dN_h(x)\over dx}\,,
\label{frag1}\\
{dN_{\pi^+}\over dx}&=&{dN_{\pi^-}\over dx}=
{dN_{\pi^0}\over dx}=\left({0.97\over3}\right)
{dN_h(x)\over dx}\,.\nonumber
\end{eqnarray}
{}From the pion injection spectra one gets the resulting
contribution to the injection spectra for $\gamma$-rays, electrons and
neutrinos.

Consider now the dependence of the decay rate of X-particles on time.
In a wide class of models involving topological defects the rate of
production (and, therefore, the rate of decay) of X-particles can be
parameterized as
\begin{equation}
  {dn_X\over dt}\propto t^{-4+p}\,,\label{Xrate}
\end{equation}
where the index $p\geq0$ depends on the nature of
topological defects from which X-particles are produced.  For
example, production of X-particles by a network of
ordinary cosmic strings in the scaling regime would correspond to
$p=1$ if one assumes that a constant fraction of the total energy of
closed loops goes into X-particles~\cite{Bh0,BR}. Models based
on annihilation of magnetic monopoles and
antimonopoles~\cite{Hill,BS} predict $p=1$ in the matter dominated and
$p=3/2$ in the radiation dominated era~\cite{SJSB} whereas simplest
models involving superconducting cosmic strings lead to
$p=0$~\cite{HSW}. A constant comoving injection rate corresponds to
$p=2$ and $p=5/2$ during the matter and radiation dominated era,
respectively.

Another model that we will consider involves primordial X-particles
with the mass of order $10^{22}$~eV and the lifetime of the order of
the age of the Universe or greater. These particles may be produced
thermally during reheating epoch~\cite{KuzRub} or from vacuum
fluctuations during inflation \cite{KuzTkach} and may constitute a
considerable fraction of Cold Dark Matter (CDM). X particle density in
this model is
\begin{equation} \label{nxkuz}
n_{X}(t)=n_{i}\frac{(1+z)^{3}}{(1+z_i)^{3}}e^{-\frac{t-t_i}
{\tau_{X}}},
\end{equation}
where $n_i$ is X-particle density at some arbitrary initial moment
$t_{i,X}$, $z$ is the redshift corresponding to time t, $z_{i,X}$ is
the initial redshift, and $10^{10}\mbox{years}\le\tau_{X}\le
10^{22}\mbox{years}$ is X-particle lifetime. We normalize X-particle
concentration by assuming that at $z_i\sim100-1000$ (matter dominated
phase) the density was approximately equal to the critical density at
that time.
Then we get
\begin{equation}
n_{i}=\frac{\rho_{0}}{m_X}(1+z_{i})^{3},
\end{equation}
where $\rho_0$ is today's critical density. Taking into account that
$t_i\ll t $ (the major contribution to the observable flux comes from
$z\la 10$) we obtain
\begin{equation}
n_{X}(t)\simeq\frac{\rho_{0}}{m_{X}}(1+z)^{3}e^{-\frac{t}{\tau_{X}}}.
\end{equation}
Finally, X particle decay rate reads
\begin{equation}
\frac{dn_{d,X}}{dt}\equiv(1+z)^{3}\left|
\frac{d}{dt}\left(\frac{n_{X}(t)}{(1+z)^{3}}\right)\right|=
\frac{\rho_{0}}{m_{X}\tau_{X}}
(1+z)^{3}e^{-\frac{t}{\tau_{X}}}.
\end{equation}

For the full determination of the injection spectra one needs also to
specify the X-particle decay mode. In what follows we assume that
X-particle decays on $n_q$ quarks and $n_e$ electrons and all these
initial decay products share energy equally: $E_q=E_e=m_X/(n_q+n_e)$.

One remark concerning the model of primordial X-particles is in order.
One has to take into account the fact that these particles are
concentrated mostly in galaxies, and so the electron component of
injection spectra is heavily suppressed due to synchrotron radiation
in the galactic magnetic field.  For the same reason there is an
additional contribution in photon spectrum. In the case of the decay
mode $X\rightarrow n_q q$ typical electron energy is
$10^{19}-10^{20}$~eV for $m_X=10^{22}$~eV and $n_q=2, 3$.  In the
galactic magnetic field of strength $B\simeq10^{-6}$G the energy of
the synchrotron photons produced by such electrons is about
$10^{14}-10^{15}$eV. Photon attenuation length at those energies is
extremely small due to pair production on microwave background, so one
should expect that the synchrotron component is also negligible. In
the present paper we approximate this effect by simply excluding
electrons from the injection spectra.  It is worth noting that the
contribution of our Galaxy to the observable flux may be
non-negligible. In this paper we present calculation of the
extragalactic component of the observable flux only. The calculation
of the Galactic contribution requires separate detailed
calculation~\cite{kks}.

\section{ Interactions of Electrons, Photons and Nucleons
with Background Radiation and Extragalactic Magnetic Field}

The $\gamma$-rays and electrons produced by X particle decay initiate
electromagnetic (EM) cascades on low energy radiation fields such as
the cosmic microwave background (CMB).  In the absence of magnetic
field EM cascades are driven mostly by the cycle of pair production
(PP; $\gamma \gamma_b\rightarrow e^- e^+$) and inverse Compton
scattering (ICS; $e \gamma_b \rightarrow e' \gamma$).  The energy
degradation of the ``leading'' particle in this cycle is slow.  Other
EM interactions that influence the $\gamma$-ray spectrum in the energy
range of interest are triplet pair production ($e \gamma_b\rightarrow
e e^- e^+$) and double pair production ($\gamma \gamma_b\rightarrow
e^-e^+e^-e^+$), as well as synchrotron losses of electrons in the
large scale extragalactic magnetic field (EGMF).

The relevant nucleon interactions implemented in the code are pair
production by protons ($p\gamma_b\rightarrow p e^- e^+$),
photoproduction of single or multiple pions ($N\gamma_b \rightarrow N
\;n\pi$, $n\geq1$) and neutron decay.  Production of secondary
$\gamma$-rays, electrons, and neutrinos by pion decay is in general
negligible in the scenarios where injection is dominated by
$\gamma$-rays and leptons over nucleons.  Apart from these processes
described in detail in ref.~\cite{Lee}, for protons of energies
$10^{15}-10^{17}$~eV the inverse Compton scattering must be taken into
account. Its cross section is identical to that for electrons (with
the replacement of $m_e$ by $m_p$).

There are two major uncertainties in the parameters which influence
the particle transport. The first one concerns the intensity and
spectrum of URB for which there exists only an estimate above a few
MHz frequency~\cite{Clark}. We adopt the so-called ``minimal'' URB
estimate that has a low-frequency cutoff at 2 MHz~\cite{Clark}.  The
other uncertainty is the mean value of EGMF and its time evolution.
Conventionally, concerning particle propagation, EGMF is assumed to be
constant and lying in the range $10^{-12}G \la B \la 10^{-9}G$
\cite{Lee,Sigl98}.  The other possibility is to assume that the total
energy of EGMF is constant; this implies $B\sim(1+z)^{3/2}$.

We assume a flat Universe with no cosmological constant, and a Hubble
constant of $h_0=0.75$ in units of $100\;{\rm km}\;{\rm sec}^{-1}{\rm
  Mpc}^{-1}$ throughout.

\section{Solution of the Transport Equations }

An example of the transport equation for electrons which
includes for simplicity only pair production (PP) and inverse Compton
scattering (ICS) looks as follows:
\begin{eqnarray}
\frac{d}{dt} N_e (E_e,t) & = & - N_e (E_e,t) \int
d\epsilon\,n(\epsilon) \int d\mu \frac{1-\beta_e \mu}{2} \sigma_{\rm
ICS} (E_e,\epsilon,\mu) + \label{Transport} \\
& & \int dE^\prime_{e}N_e(E^\prime_{e},t) \int d\epsilon\,n(\epsilon) \int d\mu
\frac{1-\beta^\prime_{e} \mu}{2}
\frac{d\sigma_{\rm ICS}}{dE_e} (E_e;E^\prime_{e},\epsilon,\mu) + \nonumber \\
& & \int dE_{\gamma}N_{\gamma} (E_{\gamma},t) \int d\epsilon\,n(\epsilon)
\int d\mu \frac{1-\mu}{2} \frac{d\sigma_{\rm PP}}{dE_e}
(E_e;E_{\gamma},\epsilon,\mu) + Q(E_e,t),\nonumber
\end{eqnarray}
where $N_e (E_e,t)$ is the (differential) number density of electrons
with energy $E_e$ at time $t$, $n(\epsilon)$ is the number density of
background photons of energy $\epsilon$, $Q(E_e,t)$ is an external
source term for electrons of energy $E_e$ at time $t$, $\mu$ is the
cosine of the interaction angle between the CR electron and the
background photon ($\mu=-1$ for a head-on collision), and $\beta_e$ is
the velocity of the CR electron. The first term in (\ref{Transport})
describes the loss of electrons due to ICS, the second one accounts
for influx of electrons scattered into corresponding energy range due
to ICS, and the third term describes the influx of electrons produced
due to PP by photons.  The processes with continuous energy loss
$dE/dt=-f(E,t)$, such as synchrotron radiation, may be accounted for
by including into the transport equations the additional term
\begin{equation}
\frac{dN}{dt}= ...+f(E,t)\frac{dN}{dE}
\end{equation}

In order to solve the transport equations numerically, we divide each
decade of energy into $n\sim20$ equidistant logarithmic bins and
define the central value of the $i$-th bin as $E_i$. Then we replace
the continuous integrals by finite sums, and integrate
Eq.~(\ref{Transport}) over one CR energy bin.  The resulting set of
differential equations has the form
\begin{equation}
\frac{d}{dt} N_a^i  =
- N_a^i\alpha_{a}^{i} + \sum_c\sum_j \beta_{ca}^{ij}N_c^j +
 Q_a^{i}.\label{diff}
\end{equation}
where subscripts $a$ and $c$ refer to particle species. The
coefficients $\alpha_a^i$, $\beta_{ca}^{ij}$ and $Q_a^{i}$ depend on
time and may be easily obtained from the original transport equation.
For numerical solution of (\ref{diff}) we used the implicit scheme
with Richardson extrapolation \cite{nr}.

To account for redshifting S.Lee in ref. \cite{Lee} performed the
operation $N_a(E,z)\rightarrow[1+\Delta
z/(1+z)]^{-2}N_a\left(E[1+\Delta z/(1+z)],z\right)$ for each particle
species $a$ after a step $\Delta z$ in redshift. Here $\Delta z$ was
matched conveniently with the logarithmic energy bin size,
$\log_{10}[1+\Delta z/(1+z)]=\log_{10}(E_i/E_{i-1})=1/n$, which
corresponds to the transformation $N^i_a(z)\rightarrow[1+\Delta
z/(1+z)]^{-3}N^{i+1}_a(z)$.

It is important to note that the time interval $\Delta t$
corresponding to $\Delta z$ may be rather long. For example at
$z\simeq0$ and $n=20$ one has $\Delta z=0.12$, which corresponds to
distances $\Delta l\simeq360/h_0\mbox{Mpc}\sim500$~Mpc and is much
larger than the synchrotron loss length of electrons, $D\sim1$Mpc, at
$E\sim10^{23}$eV in the magnetic field $B=10^{-11}G$.  So, we have to
divide time interval into smaller steps to account for all the fast
processes such as synchrotron losses of electrons, GZK-cutoff for
nucleons, etc. This was taken into account in the code.

\section{Results} \label{results}

\begin{figure}
\begin{center}
\rotatebox{90}{\epsfig{file=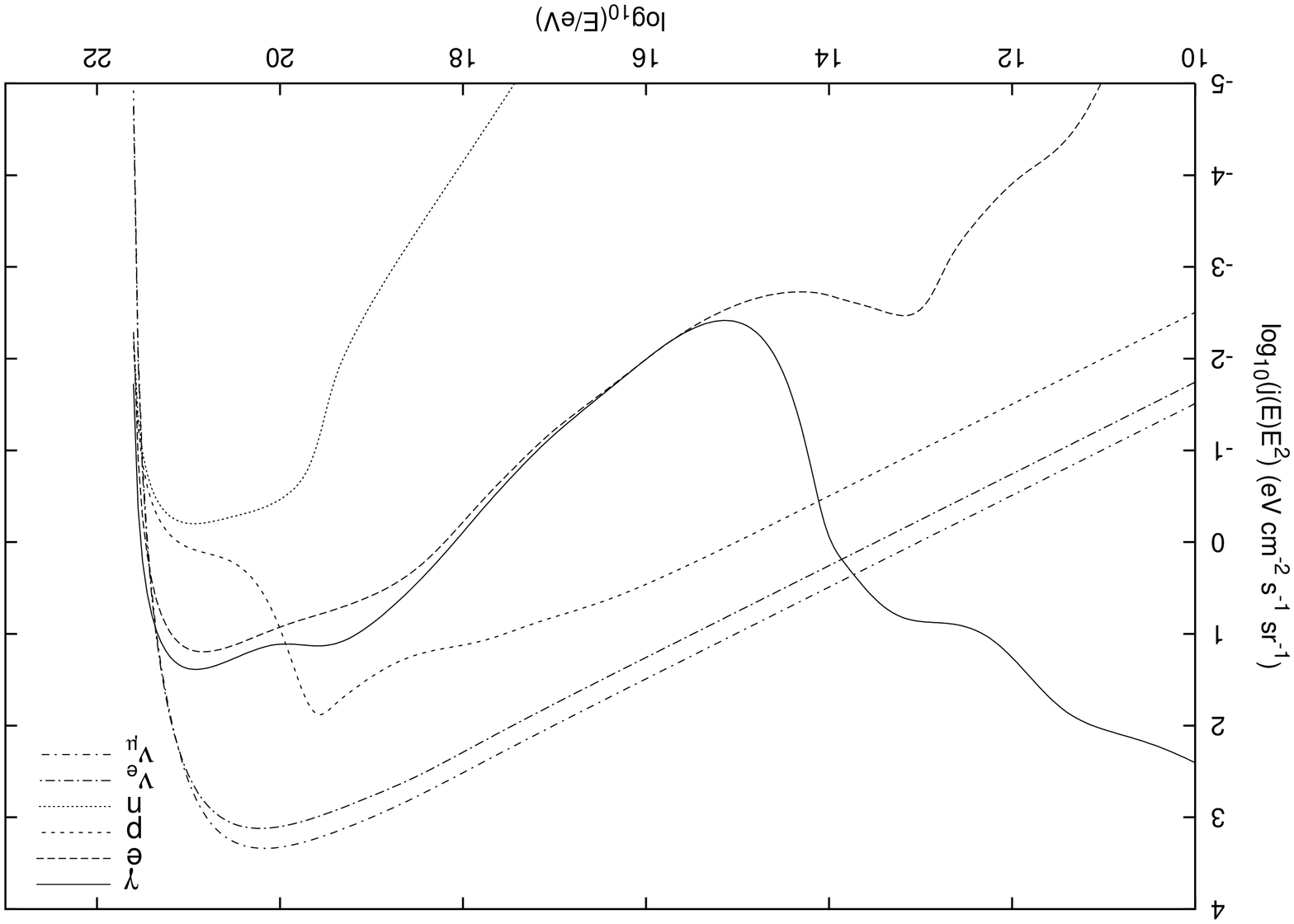,width=9cm,height=13cm}}
\end{center}
\caption{Extra-galactic component of the spectra in the model 
  of long-living X-particles with $\tau=10^9t_0$ ($t_0$ is the age of
  Universe) and the decay mode $X\rightarrow q q$. The strength of
  EGMF is $B=10^{-12}$G. Synchrotron losses in the Galaxy
  are not taken into account (see comments in Sect. \ref{results}).
\label{k9B12} }
\end{figure}

\begin{figure}
\begin{center}
\rotatebox{90}{\epsfig{file=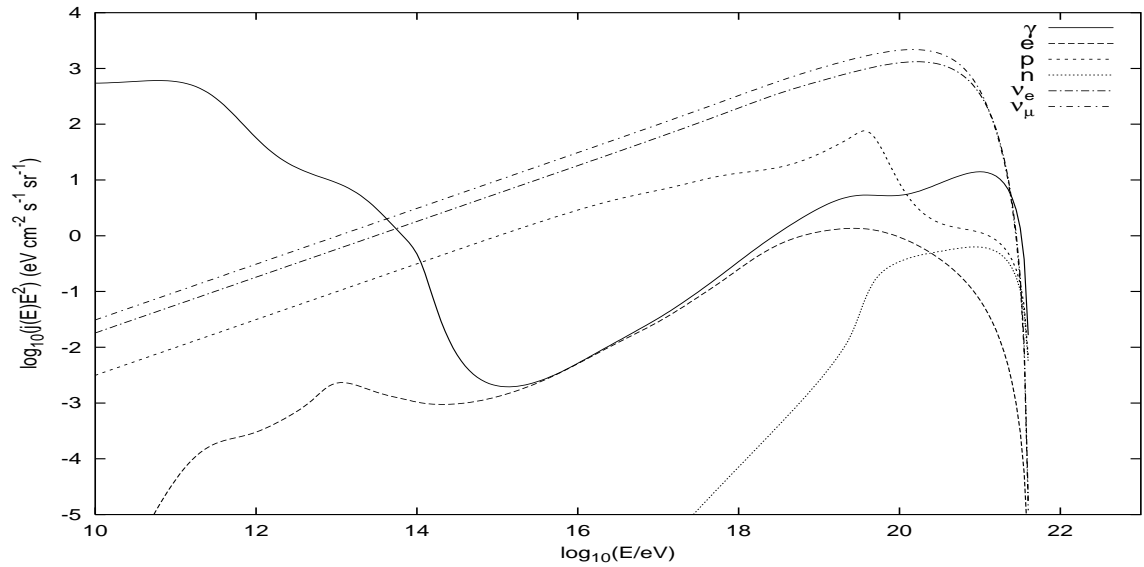,width=9cm,height=13cm}}
\end{center}
\caption{Same as fig. \ref{k9B12} but for $B=10^{-10}$G.
\label{k9B10} }
\end{figure}

\begin{figure}
\begin{center}
\rotatebox{90}{\epsfig{file=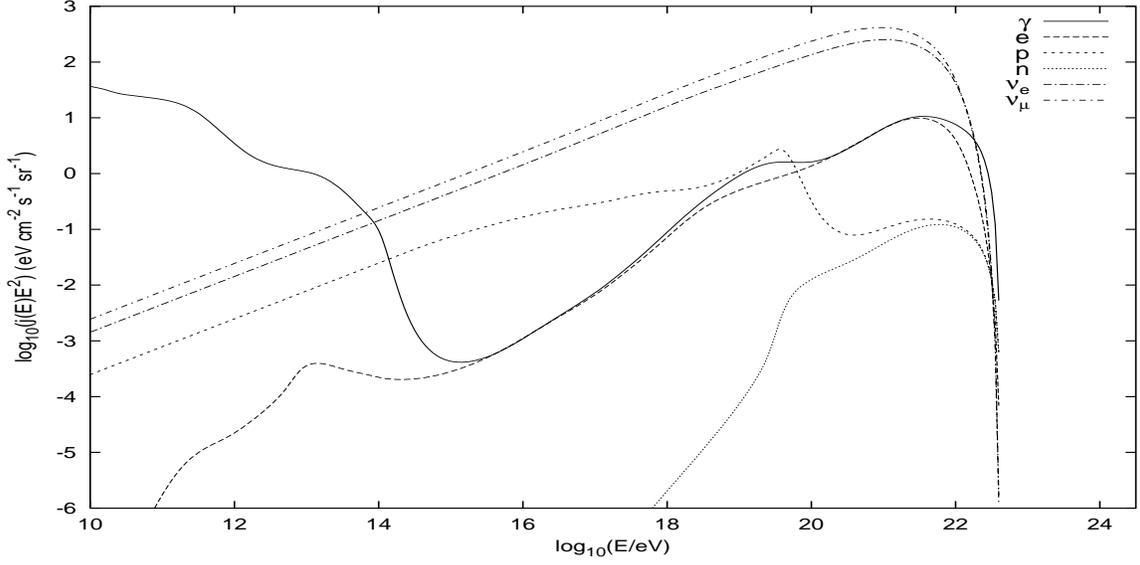,width=9cm,height=13cm}}
\end{center}
\caption{UHECR spectra near the boundary of our Galaxy 
  in the top-down model with $p=1$, $M_X=10^{23}$eV and constant EGMF
  $B=10^{-12}$G, assuming decay mode $X\rightarrow q q$. The spectra
  are normalized to the observed flux at energy $10^{20}$eV.
\label{p1qqB12} }
\end{figure}

\begin{figure}
\begin{center}
\rotatebox{90}{\epsfig{file=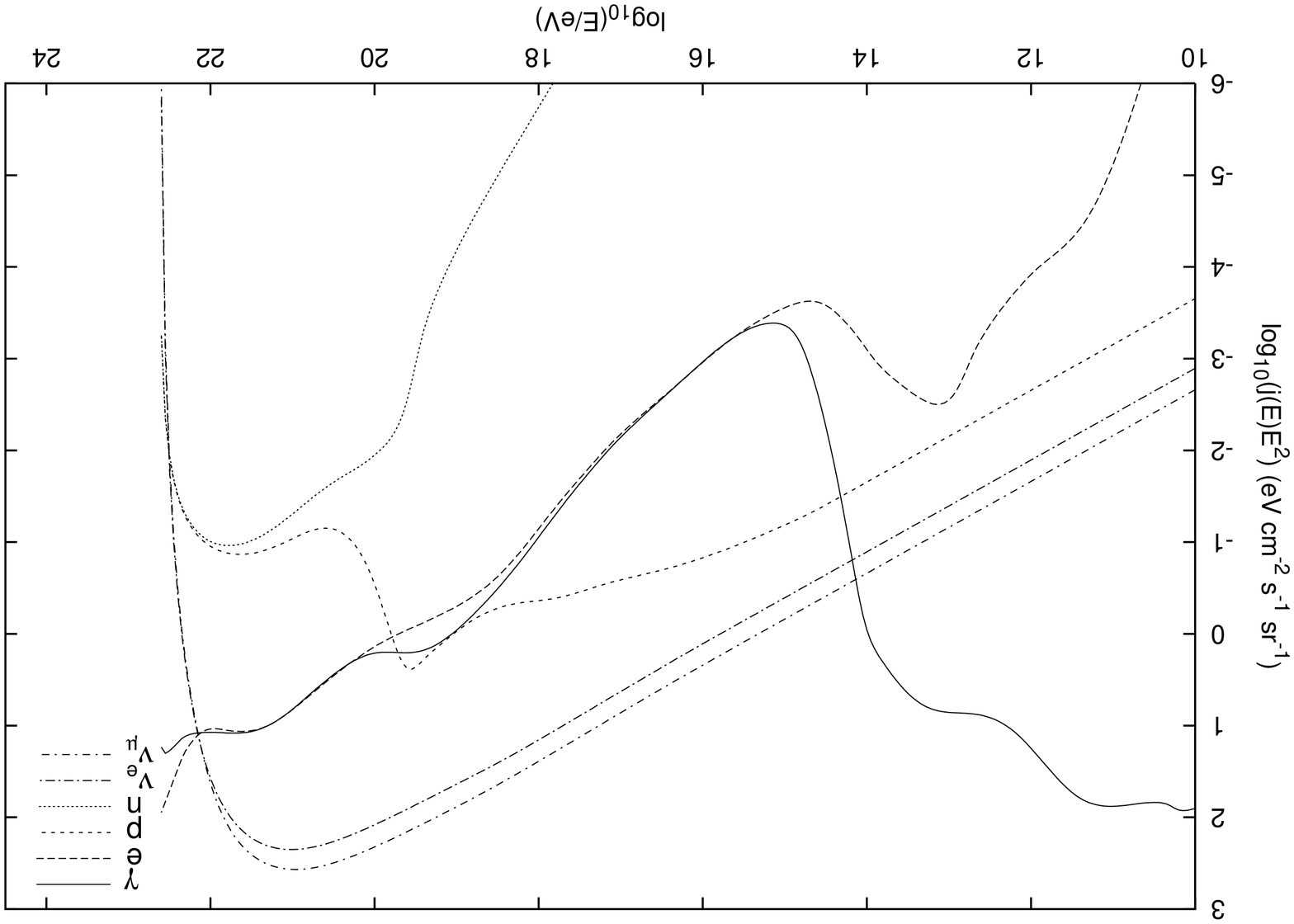,width=9cm,height=13cm}}
\end{center}
\caption{Same as fig. \ref{p1qqB12} but for the decay
mode $X\rightarrow q e$.
\label{p1qeB12} }
\end{figure}

In figs. \ref{k9B12}-\ref{p1qeB12} we present the results of our flux
calculations. Since we have not taken into account the propagation
through our Galaxy, the spectra presented refer to the fluxes of
particles near the boundary of the Galaxy. Even in the absence of
galactic sources, the observable flux of electrons will be smaller by
several orders of magnitude at energies $E \ga 10^{15}$eV, while the
photon flux will contain additional component due to the synchrotron
radiation in the galactic magnetic field \cite{Peter}.

We simulated the situation of fig. 14(a) in ref.\cite{Lee_hep}
(see fig. \ref{p1qeB12}). As a result we got the spectrum whose
shape is mostly similar to fig. 14(a) in \cite{Lee_hep}. But there are
some discrepancies in details which concerns especially the proton
component and may in part be attributed to including elastic proton-photon
scattering and also slightly different accounting for pair production
by protons.

The fluxes of UHECRs in the model of long-living X-particles are
proportional to $\tau^{-1}$, which allows to impose constraints on X
particle lifetime.  For example, in the case of decay mode
$X\rightarrow q q$ and $m_X=10^{22}$eV, comparison of the levels of
predicted spectra with the observed one leads to the constraint
\begin{equation}
\frac{\tau}{\Omega_X}\ga10^{10}t_0,
\end{equation}
where $t_0$ is the today's age of the Universe and $\Omega_X$ is the
relative density of X-particles at $z=z_i\sim100-1000$.

The main difference in spectra resulting from TD and primordial
X-particle models concerns the ratio of nucleon and electron-photon
components at energies above $10^{18}$eV. The reason is the absence of
electrons in the effective injection spectra in the second model. We
expect that this difference in spectra will persist even after
including into consideration the galactic component of the UHECR.

\section*{Acknowledgments}

The authors are grateful to P.~Tinyakov and S.~Dubovsky for many
stimulating discussions.  We also thank F.~Bezrukov, D.~Grigoriev and
G.~Pivovarov for help and interest in our work.

\newpage

\newpage

\begin{references}

\bibitem{fe1} D.~J.~Bird {\it et al.}, Phys.~Rev.~Lett.~{\bf 71}, 3401 (1993);
Astrophys.~J.~{\bf 424}, 491 (1994); {\it ibid.} {\bf 441}, 144 (1995).

\bibitem{agasa1} N.~Hayashida {\it et al.}, Phys.~Rev.~Lett.~{\bf 73}, 3491
(1994); S.~Yoshida {\it et al.}, Astropart.~Phys.~{\bf 3}, 105 (1995);
M.~Takeda {\it et al.}, Phys.~Rev.~Lett.~{\bf 81}, 1163 (1998).

\bibitem{Blandford} for a review see, e.g., R.~Blandford and
D.~Eichler, Phys. Rep. {\bf 154}, 1 (1987).

\bibitem{Hillas} A.~M.~Hillas, Ann. Rev. Astron. Astrophys. {\bf
22}, 425 (1984).

\bibitem{GZK} K.~Greisen, Phys.~Rev.~Lett.~{\bf 16}, 748 (1966);
G.~T.~Zatsepin and V.~A.~Kuzmin, Pisma Zh.~Eksp.~Teor.~Fiz.~{\bf 4}, 114
(1966) [JETP.~Lett.~{\bf 4}, 78 (1966)].

\bibitem{SSB} G.~Sigl, D.~N.~Schramm, and P.~Bhattacharjee,
Astropart. Phys. {\bf 2}, 401 (1994); J.~W.~Elbert and P.~Sommers,
Astrophys. J. {\bf 441}, 151 (1995).

\bibitem{Puget} J.~L.~Puget, F.~W.~Stecker, and J.~H.~Bredekamp,
Astrophys. J. {\bf 205}, 638 (1976); L.~N.~Epele and E.~Roulet,
e-print astro-ph/9808104; F.~W.~Stecker and M.~H.~Salamom, e-print
astro-ph/9808110.

\bibitem{KuzRub} V.~A.~Kuzmin, V.~A.~Rubakov
Ultra-High Energy Cosmic Rays: a Window to Post-Inflationary Reheating Epoch
of the Universe?
Phys.~Atom.~Nucl.~{\bf 61} (1998) 1028-1030; ~{\bf 61} (1998) 1122-1124

\bibitem{KuzTkach}
V. Kuzmin and I. Tkachov: {\it Ultra-High Energy Cosmic Rays, Superheavy
Long-Living Particles, and Matter Creation after Inflation,}
JETP Lett.~{\bf 68} (1998) 271-275

\bibitem{Lee} S.~Lee, Phys. Rev. D {\bf 58}, 043004 (1998).

\bibitem{Bh0} P.~Bhattacharjee, Phys. Rev. D {\bf 40},
3968 (1989).

\bibitem{BR} P.~Bhattacharjee and N.~C.~Rana, Phys. Lett. B
{\bf 246}, 365 (1990).

\bibitem{Sigl98} P.~Bhattacharjee and G.~Sigl
{\it Origin and Propagation of Extremely High Energy Cosmic Rays,}
(1998) astro-ph/9811011.

\bibitem{Berezinsky98} V.~Berezinsky
{\it Ultra high energy cosmic rays from cosmological relics,}
(1998) astro-ph/9811268.

\bibitem{Hill} C.~T.~Hill, Nucl. Phys. B {\bf 224}, 469
(1983).

\bibitem{BS} P.~Bhattacharjee and G.~Sigl, Phys. Rev. D {\bf
51}, 4079 (1995).

\bibitem{SJSB} G.~Sigl, K.~Jedamzik, D.~N.~Schramm, and
V.~Berezinsky, Phys. Rev. D {\bf 52}, 6682 (1995).

\bibitem{HSW} C.~T.~Hill, D.~N.~Schramm, and T.~P.~Walker,
Phys. Rev. D {\bf 36}, 1007 (1987).

\bibitem{kks} O.E. Kalashev, V.A. Kuzmin, D.V. Semikoz, in preparation.

\bibitem{bhs} P.~Bhattacharjee, C.~T.~Hill, and D.~N.~Schramm,
Phys. Rev. Lett. {\bf 69}, 567 (1992).

\bibitem{abs} F.~A.~Aharonian, P.~Bhattacharjee, and D.~N.~Schramm,
Phys. Rev. D {\bf 46}, 4188 (1992).

\bibitem{Clark} T.~A.~Clark, L.~W.~Brown, and J.~K.~Alexander, Nature {\bf
228}, 847 (1970).

\bibitem{nr} W.~H.~Press, S.~A. ~Teukolsky, W.~T.~Vetterling, B.~P.~Flannery
Numerical Recipes in C, 2nd edition, 724. Cambridge University Press (1992)

\bibitem{Peter} S.~L.~Dubovsky, P.~G. ~Tinyakov
{\it Generation of $10^{15}$ - $10^{17}$ eV photons by UHE CR in the Galactic
magnetic filed,} e-print astro-ph/9906092

\bibitem{Lee_hep} \cite{Lee} e-print astro-ph/9604098

\end{references}
\end{document}